# Phonon Signatures of Near-Room-Temperature Phase Transition in Quasi-One-Dimensional Bi$_4$I$_4$ Topological van der Waals Material

Nidhish Thiruthukkal Puthenveettil,[1,2] Topojit Debnath,[3] Clayton Mantz,[4] Zahra Ebrahim Nataj,[1,2] Jordan Teeter,[1,2] Md. Shafayat Hossain,[1,2,6] Fariborz Kargar,[5] Tina T. Salguero,[4] Roger K. Lake,[3] and Alexander A. Balandin[1,2,6,*]

[1]Department of Materials Science and Engineering, University of California, Los Angeles, California 90095 USA

[2]California NanoSystems Institute, University of California, Los Angeles, California 90095 USA

[3]Department of Electrical and Computer Engineering, University of California, Riverside, California 92521 USA

[4]Department of Chemistry, University of Georgia, Athens, Georgia 30602 USA

[5]Materials Research and Education Center, Department of Mechanical Engineering, Auburn University, Auburn, Alabama 36849 USA

[6]Center for Quantum Science and Engineering, University of California, Los Angeles, California 90095 USA

---

[*] Corresponding author: balandin@seas.ucla.edu

# ABSTRACT

The quasi-one-dimensional material Bi$_4$I$_4$ hosts two crystallographically similar polymorphs that realize distinct topological insulating phases separated by a first-order structural transition near room temperature. This α–β transition occurs without a change in space group, arising instead from a subtle rearrangement of chain stacking registry. Polarization-resolved Raman spectroscopy directly resolves this structural-topological phase transition through abrupt, hysteretic modifications of the phonon spectrum. Angle-dependent measurements establish the symmetry of the dominant Raman-active modes and require a complex Raman tensor formalism to account for absorption-induced phase effects. Across the transition, selected $A_g$ modes exhibit discontinuous, reversible shifts in frequency, linewidth, and relative intensity despite the absence of a space-group change. Density functional theory calculations reproduce the direction of the observed phonon renormalizations and confirm their sensitivity to stacking-dependent force constants. These results demonstrate that polarization-resolved Raman spectroscopy can detect subtle stacking-driven structural rearrangements that underlie topological band character, even when global crystallographic symmetry remains unchanged. The obtained results provide valuable insights into the interplay among lattice dynamics, structural distortions, and topological properties in this class of low-dimensional materials, with strong potential for unique functionalities.

1. Introduction

Topological quantum materials have emerged as a major platform for next-generation electronic and photonic technologies owing to their symmetry-protected electronic states and robust transport properties[1–4]. Their topologically nontrivial band structures lead to unconventional electronic and optical responses, including symmetry-protected boundary states, spin-momentum locking, and robust surface conduction. These features make these materials promising platforms for future spintronic, low-power electronic, and broadband optoelectronic technologies.[5–10] The latter motivates significant efforts to identify and engineer candidates with tunable topological phases and controllable phase transformations[11–14]. However, the realization of functional devices ultimately requires reliable methods to detect, characterize, and track the evolution of topological phases, particularly under realistic operating conditions[15]. This challenge is especially acute in systems where topological changes occur without obvious symmetry breaking. Spectroscopic probes offer a powerful route to address this problem by providing access to the microscopic processes that accompany phase transformations[16–18]. Furthermore, materials that exhibit reversible structural transformations coupled to changes in band topology provide attractive platforms for exploring how lattice degrees of freedom influence electronic topology[19–21]. Establishing clear and robust spectroscopic signatures of topological transformations is thus essential for both the fundamental understanding and practical implementation of topological functional materials.

Topological phases are governed by electronic band topology and crystal symmetry, both of which are highly sensitive to structural rearrangements[22–25]. Because lattice distortions modify interatomic bonding, orbital overlap, and symmetry relationships, phonon dynamics can provide a direct window into the microscopic mechanisms underlying topological phase transitions[26,27]. In systems with distinct structural polymorphs that host different topological states, spectroscopic probes of lattice vibrations become particularly powerful. Quasi-one-dimensional (1D) van der Waals materials are especially compelling in this context, as small relative shifts between adjacent chains can drive significant changes in electronic topology[20,21,24,28,29]. $Bi_4I_4$ is a prime example of such behavior. It crystallizes in two polymorphs with distinct topological character: a low-temperature α-phase, identified as a higher-order topological insulator (HOTI) hosting helical hinge modes[20], and a high-temperature β-phase stable above ~300 K[21,24,28–33], which has been predicted to be non-trivial as a weak topological insulator (WTI) [20]. ARPES measurements have revealed that β-Bi$_4$I$_4$ exhibits gapless Dirac surface states on the (100) surface with a bulk gap of ~100 meV at $\bar{M}$, whereas the α-phase is fully gapped on both surfaces, with gaps of ~85 meV on the (001) surface and ~35 meV on the (100) surface[20]. Based on these data, one can consider Bi$_4$I$_4$ a topological narrow-gap semiconductor.

Both phases crystallize in the monoclinic space group *C2/m* and feature 1D chains, four bismuth atoms in cross-section, running along the *b*-axis, surrounded by iodine atoms[20,28]. The difference between these polymorphs originates in a slight shift in the registry of the chains. The α-phase contains two offset Bi$_4$I$_4$ chains per unit cell, creating a staggered stacking arrangement, whereas the β-phase adopts an aligned stacking arrangement (Supplementary Figure S1)[20,28]. The key difference in chain geometry is a relative shift between adjacent Bi$_4$I$_4$ chains along the *a*-axis. In the α-phase, this shift induces hybridization between neighboring 2D quantum spin Hall (QSH) edge states, opening a gap in the (100) surface spectrum so that only hinge states remain[20,21,33]. In contrast, the β-phase retains surface states consistent with a stacking of 2D QSH layers that preserves the requisite symmetry, yielding the WTI phase[21]. Owing to a doubling of the chain stacking along *c*, the α-Bi$_4$I$_4$ unit cell is nearly twice as long as that of β-Bi$_4$I$_4$ along the *c* direction, as shown in Figures 1 (a, b)[21].

The α–β transition in Bi$_4$I$_4$ is first order and occurs near room temperature, providing an experimentally convenient opportunity to investigate how small structural rearrangements can induce changes in topological band character without altering global crystallographic symmetry [20,21]. In particular, direct experimental probes of the lattice dynamics associated with this transition, and their connection to topology, remain limited. Additionally, it is unclear whether stacking-driven structural changes that preserve space group symmetry can be reliably detected through vibrational spectroscopy. Here we address these questions using polarization-resolved Raman spectroscopy to probe the α–β phase transition in Bi$_4$I$_4$.

## 2. Results and Discussion

**2.1 Material Preparation and Characterization**

We synthesized $Bi_4I_4$ crystals using chemical vapor transport (CVT), which yielded shiny, silver-colored crystals of β-$Bi_4I_4$ having millimeter lengths, widths of up to 0.3 mm, and thicknesses approaching 0.1 mm, as shown in Figure 1 (c). $Bi_4I_4$ crystals were exfoliated by probe sonication in ethanol, yielding samples suitable for electron microscopy as shown in Figure 1 (d). Elemental composition was assessed by scanning electron microscopy-energy dispersive spectroscopy (SEM-EDS). As presented in Supplementary Figure S4, the measured atomic percentages of bismuth and iodine from mechanically exfoliated crystals were consistent within experimental uncertainty with the expected 1:1 stoichiometry of $Bi_4I_4$. To validate the phase transition temperature of the sample and confirm the thermodynamic nature of the phase change, differential scanning calorimetry (DSC) was performed. The DSC measurements provide insights into changes in the material's physical properties as a function of temperature. Figure 1 (e) presents the DSC data for both heating and cooling cycles, showing well-defined peaks corresponding to the topological phase transition at ~303 K on heating and ~297 K on cooling, respectively. As independent structural confirmations of the two polymorphs, both single-crystal X-ray diffraction and high-resolution transmission electron microscopy (HRTEM) were performed at temperatures below and above the transition, providing direct confirmation of the $Bi_4I_4$ structures. The refined crystal structures were in good agreement with published reports. Furthermore, as shown in Figures 1 (f, g), atomic resolution images acquired at temperatures below and above 300 K resolve the characteristic chains of bismuth atoms and rows of iodine atoms consistent with α-$Bi_4I_4$ and β-$Bi_4I_4$, respectively.

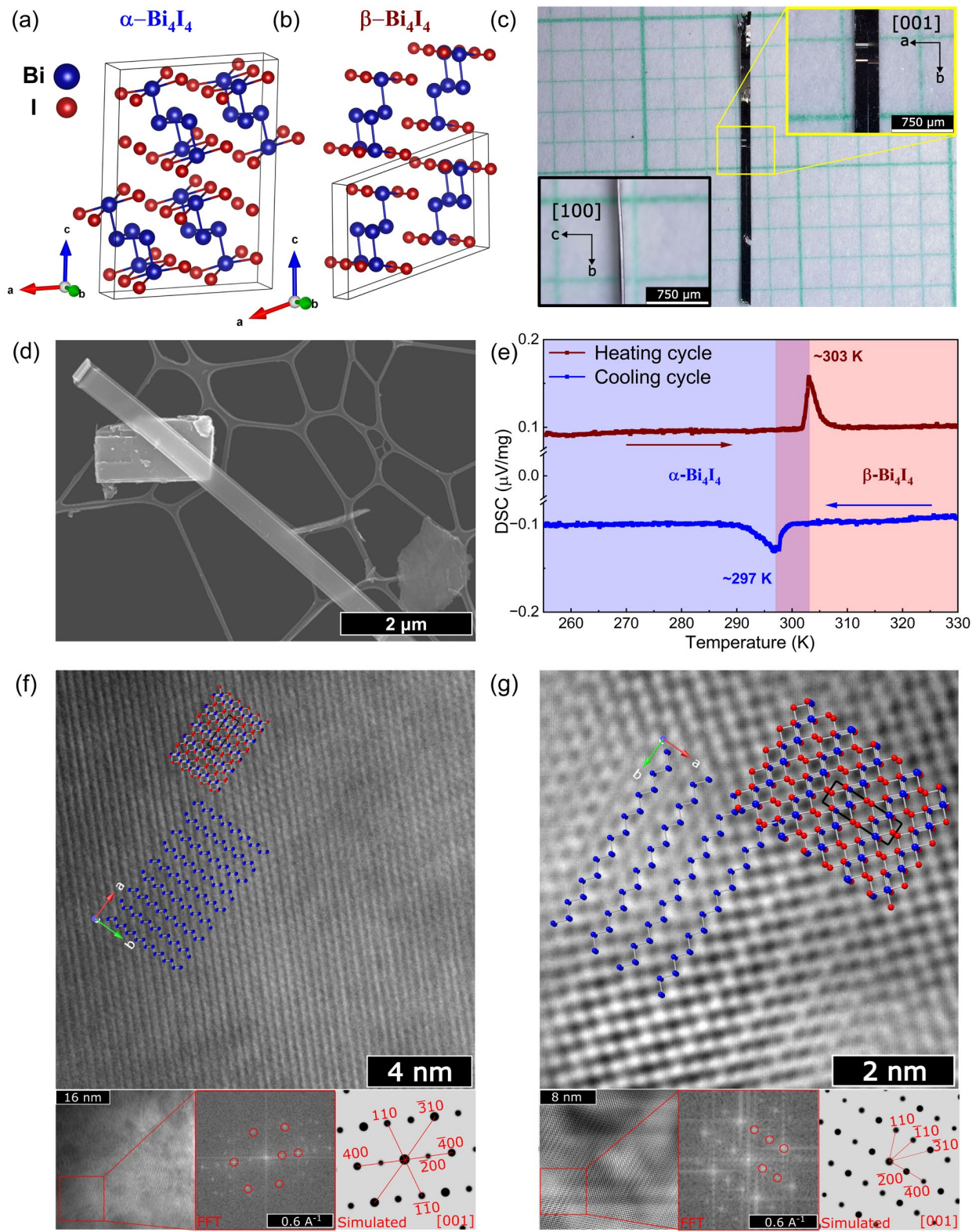

*Fig. 1: Crystal structure and characterization of Bi$_4$I$_4$. (a, b) Crystal structures of α-Bi$_4$I$_4$ and β-Bi$_4$I$_4$, respectively. The dark lines indicate the unit cells of each phase. The blue and red spheres*

*indicate Bi and I atoms, respectively. (c) Light microscopy images of an as-grown Bi$_4$I$_4$ crystal on 1 mm$^2$ graph paper. (d) Secondary electron image of exfoliated Bi$_4$I$_4$. (e) Differential scanning calorimetry (DSC) measured during cooling (blue) and heating (red) cycles, revealing a first-order structural phase transition between the α- and β-phases, with transition temperatures at ~297 K (cooling) and ~303 K (heating), indicative of thermal hysteresis. (f, g) High-resolution TEM images of (f) α-Bi$_4$I$_4$ and (g) β-Bi$_4$I$_4$, with the corresponding crystal structures overlaid to highlight the atomic arrangements in each phase. The lower panels (left to right) show the selected areas of analysis, the fast Fourier transforms (FFT), and simulated patterns for the [001] projection.]*

Following structural and thermodynamic verification of the phase transition, we performed Raman spectroscopy on bulk Bi$_4$I$_4$ crystals. Bi$_4$I$_4$ crystallizes in the monoclinic space group *C2/m* in both the α- and β-phases[20]. The α-phase contains 16 atoms per primitive cell and has 3 acoustic phonon branches and 45 optical phonon branches. The β-phase contains 8 atoms per primitive cell and has 3 acoustic phonon branches and 21 optical phonon branches. The irreducible representations of the Γ-point Raman-active modes are $A_g$ and $B_g$, which are accessible depending on the incident and scattered polarization configuration[34]. To probe the symmetry of these modes, polarization-dependent Raman measurements were performed by rotating the incident polarization within the sample plane. The azimuthal angle $\theta$ is defined as the in-plane angle between the electric field vector of the incident light and the crystallographic *b*-axis (chain direction).

## 2.2 Polarization-Dependent and Angle-Resolved Raman Spectroscopy

Figures 2 (a, d) show the room-temperature Raman spectra of Bi$_4$I$_4$ acquired with the incident laser polarization at an azimuthal angle $\theta$ = 90°, corresponding to the incident polarization aligned along the *a*-axis. The scattered light was analyzed in parallel, cross, and unpolarized configurations using 633 nm and 488 nm excitation, respectively. The incident power on the sample was maintained below 1 mW for all measurements. The Raman peaks were fitted using Lorentzian line shapes in both parallel and cross polarization configurations to extract peak positions and intensities. The initial mode assignments are based on polarization selection rules for azimuthal angle $\theta$ = 90° [35,36]. Modes that remain strong in the cross-polarized geometry but are strongly suppressed in the parallel geometry are identified as $B_g$ modes, whereas modes that are strong in the parallel geometry and strongly suppressed in the cross-polarized geometry are identified as $A_g$ modes. To establish the polarization selection rules and assign vibrational symmetries, angle-dependent measurements were performed with the sample mounted on a manual Thorlabs rotation stage with 2° graduations. Figures 2 (b, c) present the angular intensity maps for the cross- and parallel-polarized configurations under 633 nm excitation. The corresponding angular intensity maps obtained with 488 nm excitation are shown in Figures 2 (e, f). The detailed confirmation of these assignments is presented in the subsequent analysis. For the construction of the angular intensity maps, the Raman intensity of each mode was independently normalized to its maximum observed value to facilitate comparison of polarization-dependent trends. The vibrational mode at ~84 cm$^{-1}$ was not analyzed in the parallel-polarized configuration due to its very low signal-to-noise ratio.

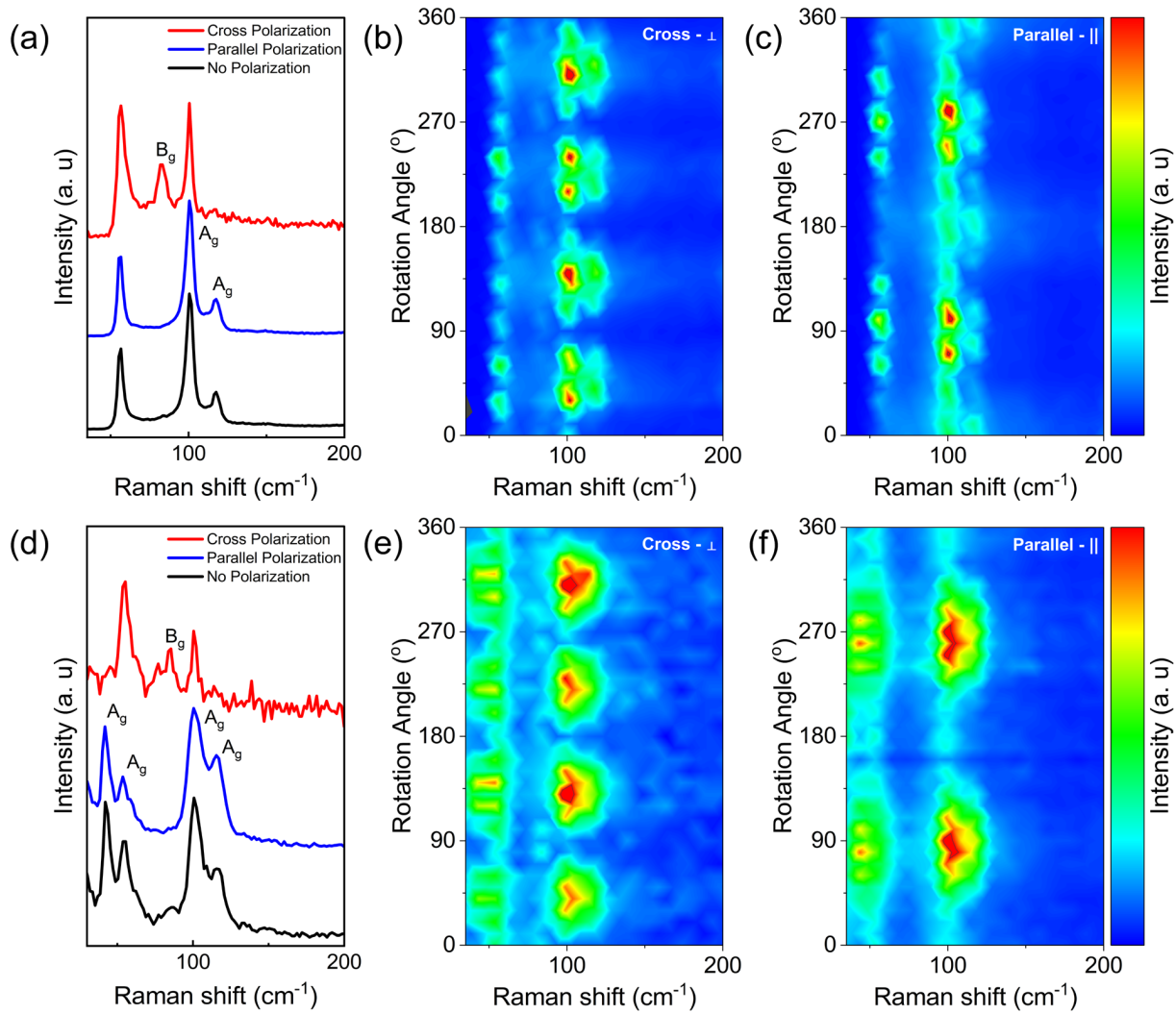

*[Fig. 2: Polarization-dependent and angle-resolved Raman spectroscopy of $Bi_4I_4$. (a) Polarization-dependent Raman spectra measured with the incident light polarization oriented parallel to the crystallographic a-axis using 633 nm (red) laser excitation. The instrumental cutoff for this laser is ~50 cm$^{-1}$. Spectra are shown for cross-, parallel-, and unpolarized in the backscattering geometries, revealing clear polarization selectivity of the Raman-active modes. The feature ~55 cm$^{-1}$ lies close to the cutoff and is therefore not included in the symmetry assignment. (b, c) Angle-resolved Raman intensity maps acquired with 633 nm excitation in the (b) cross-, and (c) parallel-polarization geometries, showing the angular modulation of Raman-active modes as a function of in-plane rotation. (d) Corresponding polarization-dependent spectra, and angle-resolved intensity maps in (e) cross-, and (f) parallel-polarization geometries measured using 488* *nm excitation. In this configuration, the instrumental cutoff is extended down to 10 cm$^{-1}$, enabling additional Raman features to be resolved at smaller Raman shifts.]*

All Raman measurements were performed on the cleaved (001) surface of Bi$_4$I$_4$[20]. Because Bi$_4$I$_4$ crystallizes in a monoclinic structure where the lattice vectors *a*, *b*, and *c* are not mutually orthogonal, we introduce a laboratory Cartesian coordinate system (X, Y, Z) to clearly define the scattering geometry and polarization directions. The azimuthal angle $\theta$ is measured about the surface normal (laboratory X axis), from the laboratory Y axis to the crystallographic *b*-axis. At $\theta$ = 0°, the crystallographic axes are aligned such that *b* ∥ Y and *a* ∥ Z. Any in-plane misalignment between the *b*-axis and the laboratory Y axis would introduce a constant angular offset. Since no measurable offset was observed, no rotational correction was included in the Raman tensor analysis[37]. Owing to the monoclinic structure, the surface normal (laboratory X) is not strictly parallel to the real-space *c*-axis. The laser propagates nominally along the X direction (normal to the (001) surface), while the incident and analyzed polarizations lie in the Y–Z plane, as shown in Figure 3 (a). To understand the angle-dependent scattering response and confirm the Raman vibrational symmetries, the intensities of the peaks at ~45 cm$^{-1}$, 55 cm$^{-1}$, 84 cm$^{-1}$, 100 cm$^{-1}$, and 115 cm$^{-1}$ were analyzed from spectra obtained at azimuthal angles ranging from 0° to 360° in steps of 10°. The intensity of Raman scattering according to classical Placzek approximation is given by $I \propto |\vec{e}^s \cdot \boldsymbol{R} \cdot \vec{e}^i|^2$ where $\vec{e}^i$ and $\vec{e}^s$ are unit polarization vectors of the incident and scattered light, respectively, and $\boldsymbol{R}$ is the Raman tensor of the mode in the monoclinic space group *C2/m*[38–40].

Following the Porto notations, $X(ZZ)X'$ (parallel) and $X(ZY)X'$ (cross), where $X$ and $X'$ denote the incident and scattered propagation directions, respectively, we define the parallel configuration of the incident and scattered light polarization vectors as $\vec{e}^i = (0 \quad cos\theta \quad sin\theta)^T$, and $\vec{e}^S = (0 \quad cos\theta \quad sin\theta)$, respectively. For the cross configuration, the scattered light polarization changes to $\vec{e}^S = (0 \quad sin\theta \quad -cos\theta)$[37,41,42]. Using the polarization vectors defined above, we derived the angular dependences for the $A_g$ and $B_g$ symmetries in parallel and cross configurations. The resulting expressions are summarized in Table 1[36,43–45]. Figures 3 and 4 show the angle dependence of the mode intensities under cross- and parallel-polarized configurations measured using 633 nm and 488 nm excitation, respectively. For each mode, the angle-dependent Raman intensity was independently normalized to its maximum value. Notably, the experimental results (solid spheres) deviate from the predictions of the conventional Raman tensor model (see Supplementary Figures S5 and S6).

To address this discrepancy, the influence of light absorption on the Raman tensor elements is incorporated by treating the relevant tensor components as complex quantities[46]. Here, only the tensor elements contributing to the scattering intensity in the selected geometries are shown, although in general, all tensor elements acquire complex phases:

$$b = |b|e^{i\phi_b}, c = |c|e^{i\phi_c}, f = |f|e^{i\phi_f} \qquad (1)$$

The phases of the Raman tensor elements are defined as[42]:

$$\phi_b = arctg\left[\frac{\frac{\partial \epsilon''_{YY}}{\partial q^{A_g}}}{\frac{\partial \epsilon'_{YY}}{\partial q^{A_g}}}\right], \phi_c = arctg\left[\frac{\frac{\partial \epsilon''_{ZZ}}{\partial q^{A_g}}}{\frac{\partial \epsilon'_{ZZ}}{\partial q^{A_g}}}\right], \phi_f = arctg\left[\frac{\frac{\partial \epsilon''_{YZ}}{\partial q^{B_g}}}{\frac{\partial \epsilon'_{YZ}}{\partial q^{B_g}}}\right] \qquad (2)$$

where $\phi$ represents the phase arising from the imaginary part of the dielectric function and therefore accounts for absorption[42]. Here, $\epsilon_{ij}$ is the dielectric tensor, with $\epsilon'_{ij}$ and $\epsilon''_{ij}$ denoting its real (dispersive) and imaginary (absorptive) components. Their derivatives with respect to the normal-mode coordinate $q^\mu$ determine the magnitude and phase of the Raman tensor elements for mode $\mu$. The Raman scattering intensities for the parallel and cross-polarization configurations using both the conventional and complex-valued Raman tensors are summarized in Table 1.

Table 1. Raman Tensors and scattering intensity for $A_g$ and $B_g$ modes of Bi$_4$I$_4$

| Raman Mode | Raman Scattering Intensity | |
|---|---|---|
| | Parallel Polarization $(\vec{e}^i \| \vec{e}^s)$ | Cross Polarization $(\vec{e}^i \perp \vec{e}^s)$ |
| | Conventional Real Raman Tensors | |
| $A_g$ | $\|bcos^2\theta + csin^2\theta\|^2$ | $\|(b-c)cos\theta sin\theta\|^2$ |
| $B_g$ | $\|f sin2\theta\|^2$ | $\|f cos2\theta\|^2$ |
| Raman Tensors with Complex-Valued Elements | | |
| $A_g$ | $(\|b\| cos^2\theta + \|c\|cos\phi_{cb} sin^2\theta)^2 + \|c\|^2 sin^2\phi_{cb} sin^4\theta$ | $[(\|b\| - \|c\|cos\phi_{cb})^2 + \|c\|^2 sin^2\phi_{cb}] sin^2\theta cos^2\theta$ |
| $B_g$ | $\|f sin2\theta\|^2$ | $\|f cos2\theta\|^2$ |

Here, $\phi_{ij}$ denotes the phase difference between Raman tensor elements, expressed as $\phi_i - \phi_j$. Figures 3 and 4 show the theoretical best-fit curves (solid lines) for each scattering geometry and vibrational symmetry, overlaid with the normalized experimental data (solid spheres) from 0° to 360°, obtained using the complex-valued Raman tensor elements for 633 nm and 488 nm laser excitations, respectively. Notably, light absorption does not affect the $B_g$ modes, since the associated phase factor cancels upon taking the square modulus of the Raman amplitude. In contrast, for the totally symmetric $A_g$ modes a relative phase difference between tensor elements survives in the intensity expression, modifying the response in the parallel-polarization configuration. In the cross geometry, this phase-dependent term does not contribute, and the angular dependence remains unchanged[42].

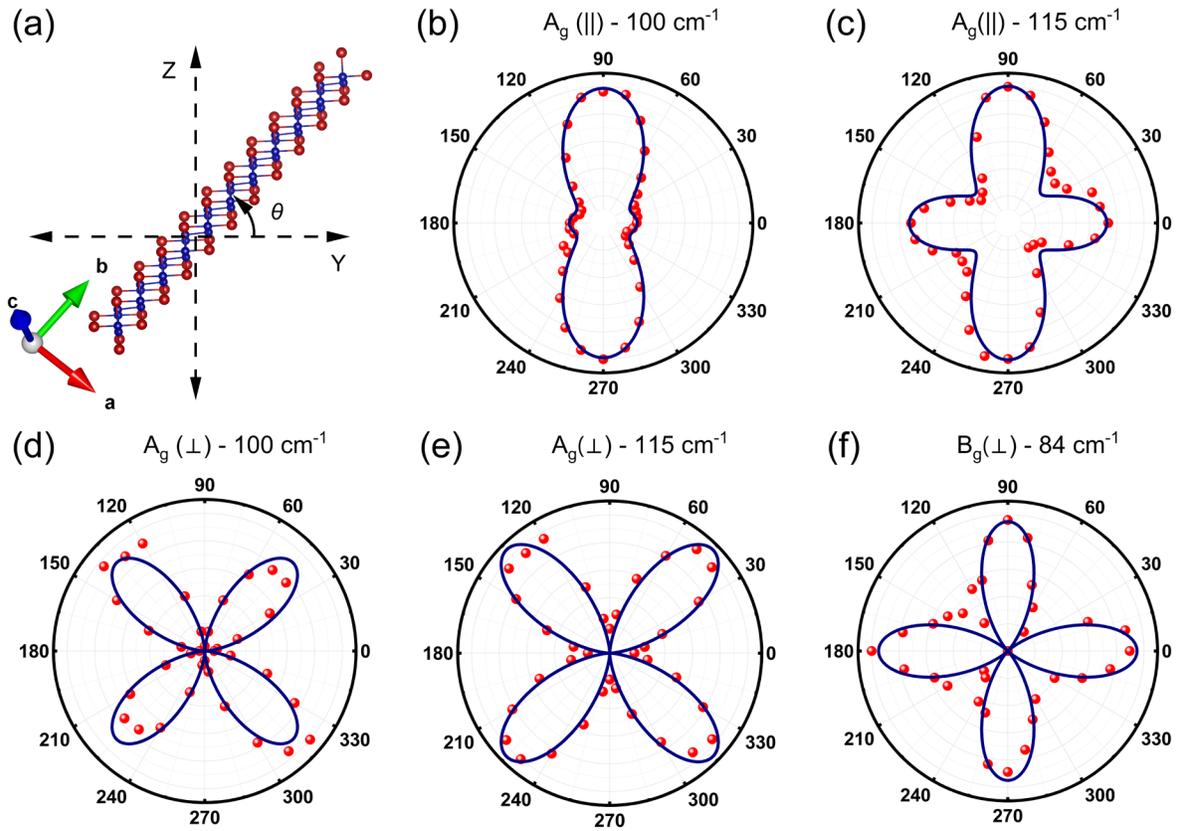

*[Fig. 3: Polar plots of normalized Raman intensities using a 633 nm laser. (a) Schematic of the angle-resolved Raman geometry for Bi$_4$I$_4$, defining the rotation angle θ between the crystallographic axis b and the lab-frame polarization (Y–Z) directions. (b, d, c, e) Polar plots of normalized intensity for $A_g$ modes at (b, d) 100 cm$^{-1}$ and (c, e) 115 cm$^{-1}$ under parallel- and cross-polarization, respectively. (f) Polar plot for $B_g$ mode at 84 cm$^{-1}$ under cross polarization configuration. The solid circles represent the experimental data, while the solid lines denote fits obtained from the intensity functions derived using Raman tensors for the C2/m space group. Parallel (∥) and cross (⊥) polarization configurations are indicated beside the symmetry and frequency of each mode.]*

The anomalous angular responses of the $A_g$ modes arise from the strong optical anisotropy and wavelength-scale absorption in Bi$_4$I$_4$, which imparts complex phase factors to the Raman tensor elements[39,42,46]. In Bi$_4$I$_4$, linear dichroism leads to unequal attenuation of polarization components along different crystallographic directions, rendering the conventional real-valued Raman tensor model insufficient for describing the fully symmetric $A_g$ vibrations[42]. Angle-dependent polarized Raman measurements enable unambiguous assignment of the vibrational symmetries of the dominant phonon modes in Bi$_4$I$_4$. Importantly, whereas earlier work attributed the feature near 100 cm$^{-1}$ to a $B_g$ mode, our Raman measurements demonstrate that this mode exhibits the characteristic angular dependence of an $A_g$ mode[31,47]. This reassignment refines the symmetry classification of the Γ-point phonons and is essential for correctly interpreting the lattice dynamics and their coupling to the underlying crystal structure.

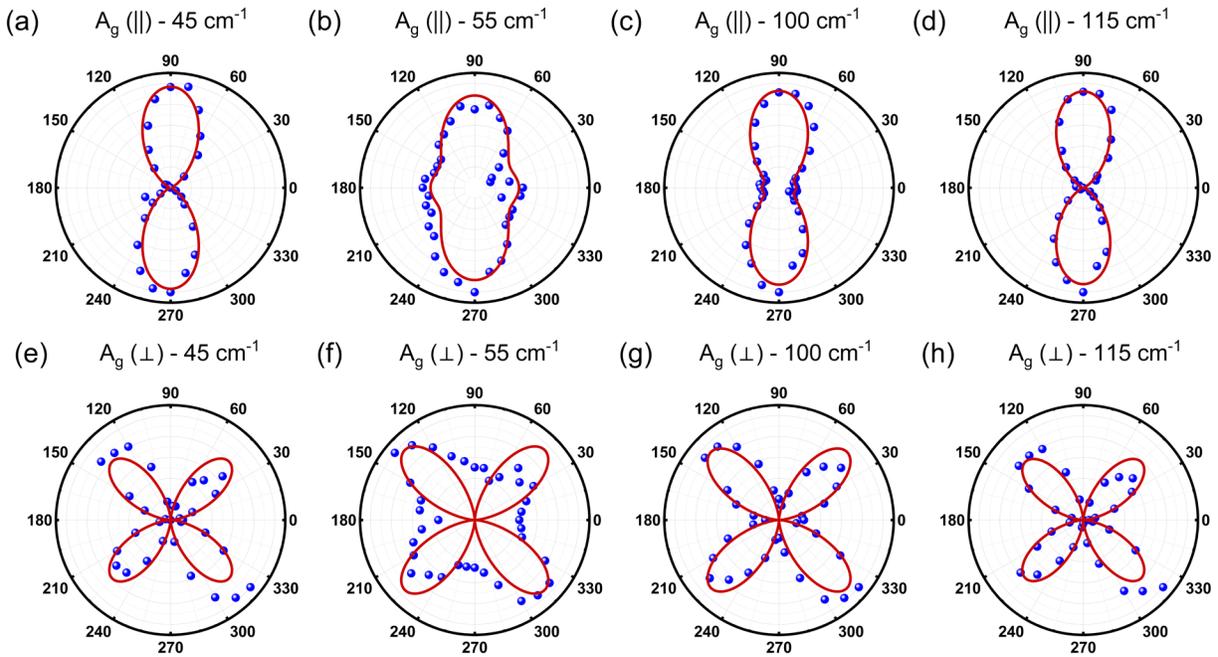

*[Fig. 4: Polar plots of normalized Raman intensities using 488 nm laser. (a-d) Polar plots of normalized Raman intensities for selected $A_g$ modes at 45, 55, 100, and 115 cm$^{-1}$, respectively, measured in the parallel polarization configuration. (e-h) Corresponding polar plots of the same $A_g$ modes measured under cross polarization configuration. ]*

## 2.3 Temperature-Dependent Raman Spectroscopy of Topological Transitions

Learning from the angle-dependent polarized Raman analysis, we performed temperature-dependent polarized Raman measurements on bulk $Bi_4I_4$ crystals from 100 K to 350 K, spanning the α–β topological phase transition near 300 K. All measurements were carried out using 488 nm excitation in the parallel polarization configuration, with the incident and analyzed polarizations aligned parallel to the crystallographic *a*-axis. This geometry maximizes the intensity of the $A_g$ modes while suppressing the weak $B_g$ contributions, as dictated by the Raman tensor selection rules derived earlier, thereby enabling precise tracking of the dominant fully symmetric phonons[28]. Figures 5 (a, b) present the parallel-polarized Raman spectra collected during cooling and heating cycles between 100 K and 350 K, respectively. The shaded spectra highlight the transition region. An anomalous spectral evolution is observed at 296 K during cooling and 304 K during heating, consistent with a thermal hysteresis of ~8 K[20,24]. These temperatures are identified as the transition points based on discontinuous shifts in phonon frequencies. The temperature evolution of the $A_g$ phonon frequencies is summarized in Figures 5 (c-f). The $A_g$ mode at 45 cm$^{-1}$ exhibits a sudden redshift of ~2 cm$^{-1}$ upon transitioning from the α-phase to the β-phase. In contrast, the $A_g$ modes at 55 cm$^{-1}$, 100 cm$^{-1}$, and 115 cm$^{-1}$ display abrupt blueshifts of ~1.5–2 cm$^{-1}$ across the same transition, corresponding to a relative change of ~1–3%, which exceeds the smooth anharmonic temperature dependence observed away from the transition. The phonon frequency discontinuities are fully reversible during the cooling cycle, confirming the reproducibility of the structural transformation. The transition temperatures extracted from Raman measurements agree closely with our DSC results and are consistent with previously reported values for $Bi_4I_4$[20,24]. The clear thermal hysteresis observed in the phonon frequencies provides compelling evidence of a first-order α to β transition[20,21,48,49].

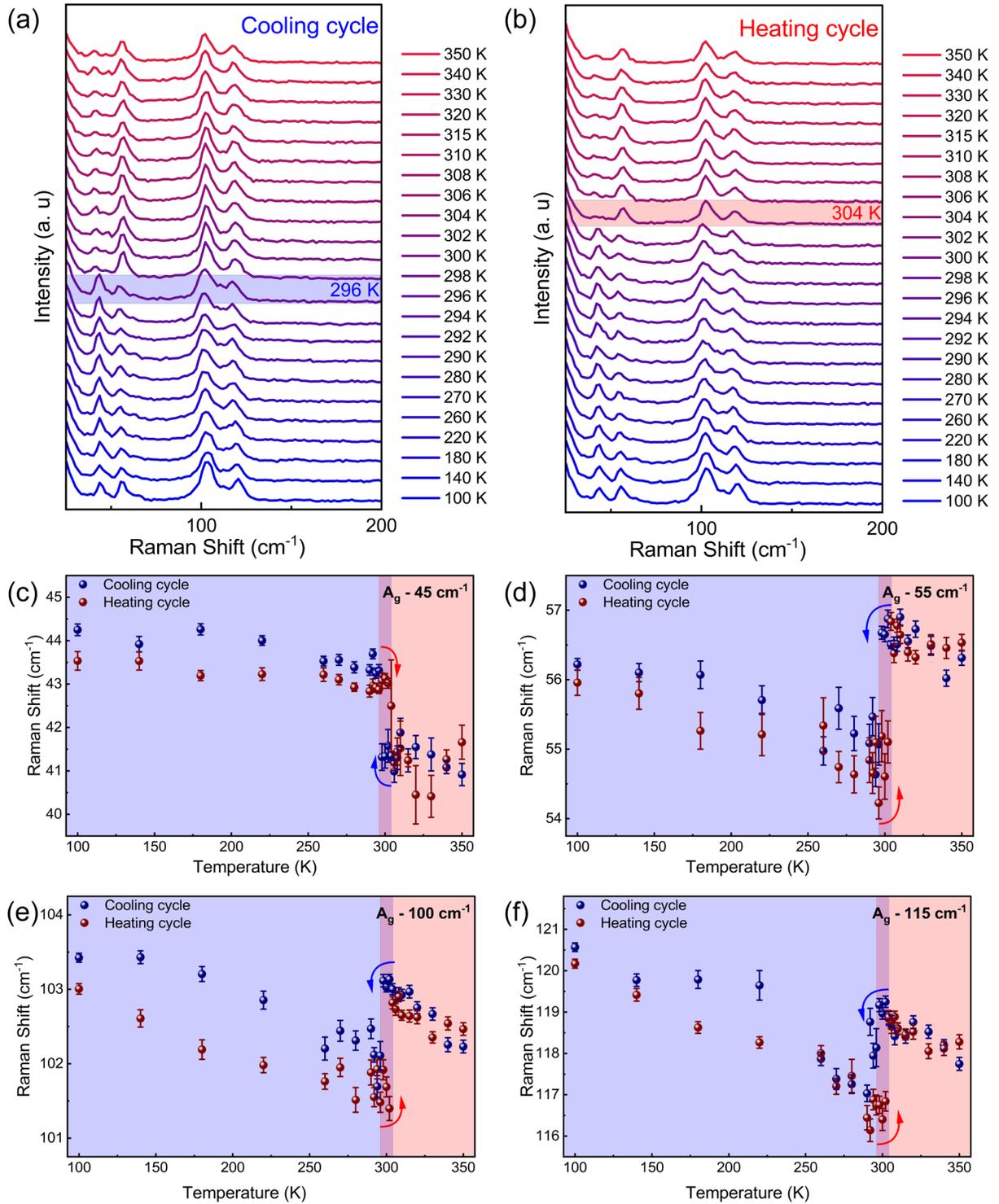

*Fig. 5: Temperature-dependent Raman signatures of the α–β topological phase transition in Bi$_4$I$_4$. (a, b)* Parallel polarized temperature-dependent Raman spectra with the shaded region indicating the phase transition temperatures measured during cooling and heating cycles with a *488 nm excitation. (c-f) Temperature evolution of the Raman peak positions for the $A_g$ modes at 45, 55, 100, and 115 cm$^{-1}$, respectively, measured under parallel polarization. Blue- and red-shaded regions indicate the stability ranges of the α- and β-phases, respectively, while the narrow vertical shaded band denotes the hysteresis window associated with the phase transition. The arrows indicate the direction of temperature sweep. Error bars represent the uncertainties obtained from Lorentzian line-shape fitting of individual Raman peaks.]*

Following the analysis of phonon frequency evolution, we further examine changes in Raman intensity and linewidth across the phase transition. In addition to the discontinuous frequency shifts, a pronounced modification in the scattering intensity of the $A_g$ modes at ~45 and 55 cm$^{-1}$ is observed at the transition. To account for possible variations in absolute collection efficiency, the intensity was quantified using the ratio of the 45 cm$^{-1}$ and 55 cm$^{-1}$ modes. The uncertainty in the intensity ratio was estimated using standard quadratic error propagation based on the fitted peak amplitudes and their associated fitting uncertainties. Figure 6 (a) shows the temperature dependence of this integrated intensity ratio during cooling and heating cycles. A clear hysteretic behavior is observed, with a step-like change occurring at the same transition temperatures identified from the phonon frequency evolution. The increased error bars in some data points reflect fitting uncertainties associated with a nearby mode at ~47 cm$^{-1}$. While this feature appears to be extremely weak in the α-phase, it becomes comparable in intensity to the ~45 cm$^{-1}$ mode in the β-phase. The discontinuity is clearly reflected in the Raman spectra: the ~55 cm$^{-1}$ mode is significantly enhanced in the β-phase, while the ~45 cm$^{-1}$ mode is concurrently suppressed compared to the α-phase.

The phonon linewidth was evaluated by extracting the full width at half maximum (FWHM) from Lorentzian fits to the $A_g$ mode at 100 cm$^{-1}$ using a consistent fitting window and linear background subtraction across all temperatures. The temperature evolution of the linewidth is presented in Figure 6 (b). A distinct reduction in FWHM is observed upon crossing from the α-phase to the β-phase, with the hysteresis following the same thermal cycle dependence as the frequency and intensity changes. Together, the thermal hysteresis features in phonon frequency, relative intensity, and linewidth provide a coherent spectroscopic signature of the temperature-driven α↔β topological phase transition in $Bi_4I_4$.

The temperature-dependent measurements further enable phase identification of the angle-resolved Raman data acquired with different excitation wavelengths. The angular measurements performed using 488 nm excitation are consistent with the α-phase, as evidenced by the phonon frequencies and the $I_{45}/I_{55}$ intensity ratio characteristic of the α-phase established from the temperature-dependent data. The angular spectra obtained with 633 nm excitation are identified as the β-phase. This assignment is confirmed by temperature-dependent Raman measurements acquired using the same 633 nm excitation (Supplementary Figure S7), which show that the observed frequency of the ~100 cm$^{-1}$ $A_g$ mode matches the β-phase. The angular measurements under 488 nm excitation were performed at incident powers below 150 μW, while the measurements under 633 nm excitation were acquired at a substantially higher power of ~800 μW. In the room-temperature angular measurements, no cryostat has been used. Near the transition region, the higher excitation power is sufficient to generate local heating to promote stabilization of the β-phase. For this reason, a Raman spectrometer can be used not only to distinguish the two structural phases through their angular and spectral fingerprints but also to induce the phase transition. The latter attests to the power of inelastic light scattering spectroscopy for the characterization of low-dimensional materials[50–52].

The 55 cm$^{-1}$ mode shows a deviation from the complex Raman tensor fit in the cross-polarized geometry under 488 nm excitation, as shown in Figure 4 (f). Notably, this mode also exhibits one of the largest intensity changes across the α↔β transition. This correlation suggests that the anomalous angular response may be sensitive to subtle stacking-dependent modifications of the Raman tensor elements beyond the single-mode approximation used in our model. Similar mode-selective anomalies have been reported in other materials, indicating that individual phonon modes can deviate from simplified Raman tensor descriptions[53–55]. Because similar deviations are not observed for the other phonon modes, the effect is unlikely to arise from global experimental factors, such as laser-induced heating. In addition, neither the experimental spectra nor the calculated phonon dispersion indicates the presence of nearby nearly degenerate $B_g$ modes that could lead to interference effects in the angular response[56]. Excitation-dependent resonant enhancement of the Raman tensor elements or phase-sensitive renormalization of the complex

Raman tensor near the transition may modify the polarization-dependent Raman intensities in ways that are not captured within our model.

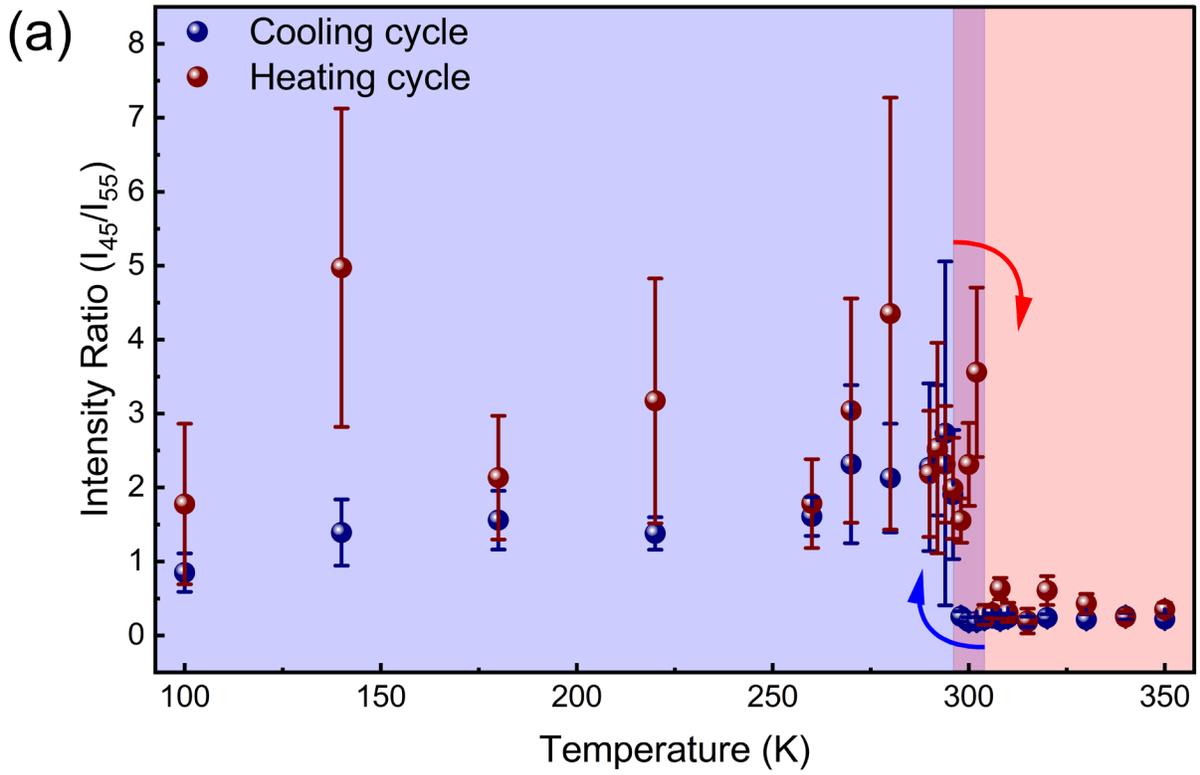

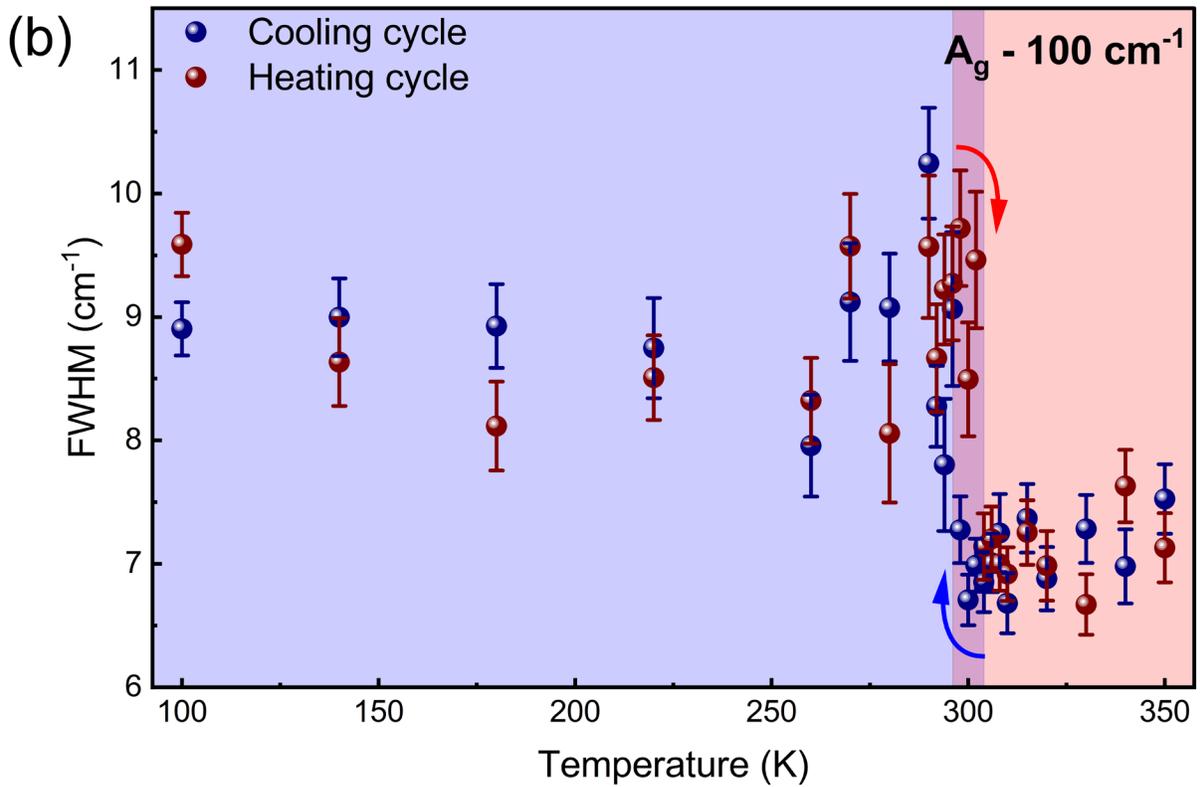

*[Fig. 6: Temperature-dependent Raman intensity ratios and linewidth evolution in $Bi_4I_4$ across*

***the phase transition*** *(a) Temperature dependence of the integrated intensity ratio between the ~45 cm$^{-1}$ and ~55 cm$^{-1}$ Raman modes ($I_{45}/I_{55}$) measured during cooling (blue symbols) and heating (red symbols) cycles. A pronounced change in the intensity ratio is observed near the α–β structural transition temperature at ~300 K. (b) Temperature evolution of the full width at half maximum (FWHM) of the $A_g$ Raman mode at 100 cm$^{-1}$ during cooling and heating cycles. An abrupt change in linewidth is observed across the transition, providing spectroscopic evidence of the phase transition.]*

To further understand the changes observed across the phase transition, we calculated the phonon dispersions of both α- and β-$Bi_4I_4$. Because the lattice parameter along the *c*-axis differs between the two phases, the Brillouin zone also changes[20]. In particular, the doubling of the unit cell along the *c*-axis reduces the bulk Brillouin zone by half along the Z direction, as shown in Figures 7 (a, b) and correspondingly increases the number of Raman-active modes through zone folding. The calculated phonon dispersions are in good overall agreement with the experimental spectra. Figures 7 (c, d) show the phonon band structures, with the experimentally observed Raman modes highlighted at the Γ-point by red and green spheres for the α- and β-phases, respectively. The experimentally observed peak frequencies and vibrational symmetries match well with the calculated values. Notably, the calculated frequencies of the modes near 55, 100, and 115 cm$^{-1}$ are higher in the β-phase than in the α-phase, consistent with the experimentally observed blueshift at the α→β transition. Conversely, the mode at ~45 cm$^{-1}$ is calculated to have a lower frequency in the β-phase, consistent with its experimentally observed redshift across the transition.

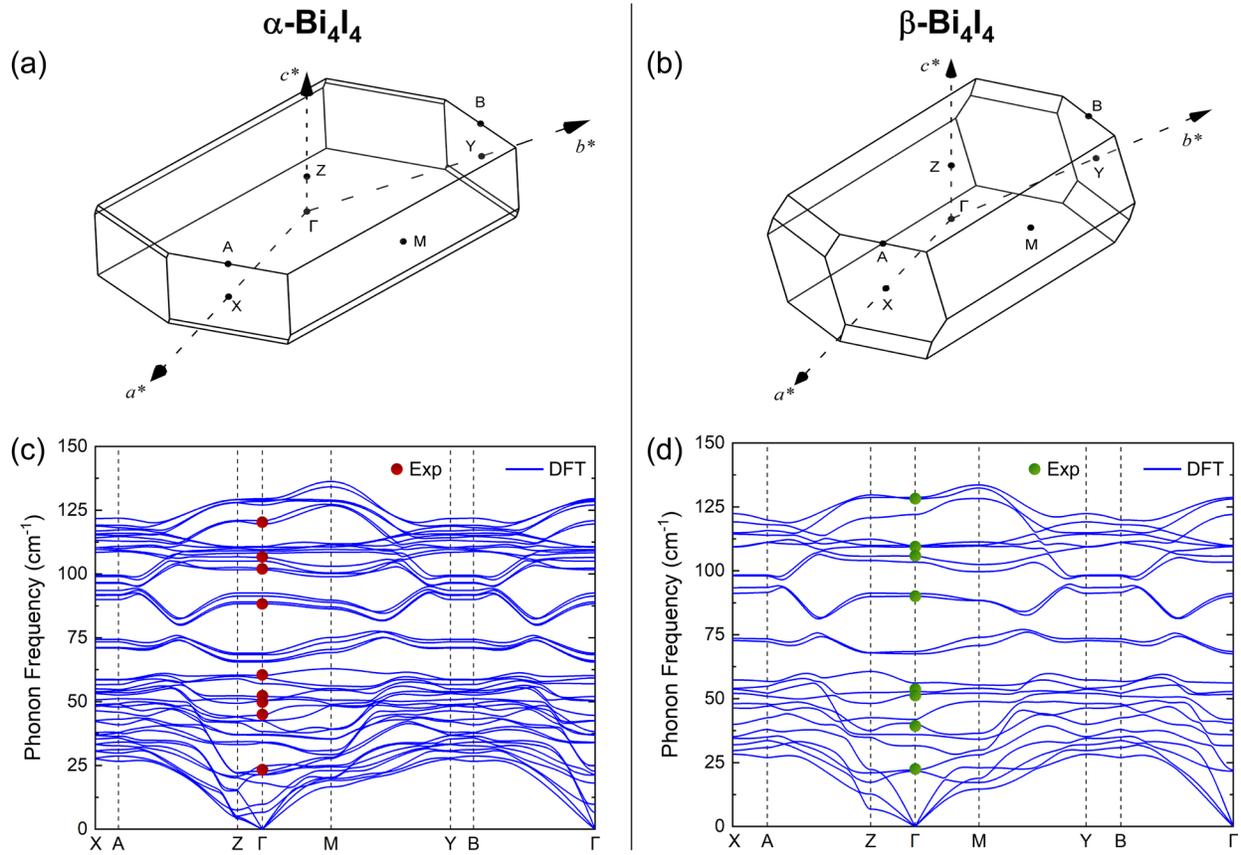

*[Fig. 7: **Calculated phonon dispersions of Bi$_4$I$_4$**. (a, b) First Brillouin zones of the (a) α- and (b) β-phases with the corresponding high-symmetry points indicated. (c, d) Calculated phonon dispersions of the (c) α- and (d) β-phases, plotted along the high-symmetry paths. Red and green spheres indicate experimentally observed phonon modes in α-Bi$_4$I$_4$ and β-Bi$_4$I$_4$, respectively.]*

While the DFT results reproduce the mode symmetries and the direction of the frequency changes, they do not by themselves identify the microscopic origin of the transition-induced anomalies. To provide further insight into the Raman changes across the transition, cleavage energies of both α- and β-Bi$_4$I$_4$ were evaluated by introducing separation across the van der Waals gaps (Supplementary Figure S8). For both α- and β-Bi$_4$I$_4$, the cleavage energy converges to approximately 0.27 J m$^{-2}$, indicating similarly weak binding between neighboring ribbon-like units in the two phases. The exfoliation energies obtained from the slab calculations are nearly identical to the corresponding cleavage energies, implying that the energetic cost of removing a terminal unit is essentially the same as that required to separate the crystal across the same weakly bonded interface. For comparison, graphite exhibits a cleavage energy of approximately 0.32 J m$^{-2}$, slightly larger than the value obtained here[57]. The relatively small energies found for both phases of $Bi_4I_4$ therefore confirm the weak van der Waals character of the inter-unit bonding and support the feasibility of mechanical thinning or exfoliation. Moreover, the close agreement between cleavage and exfoliation energies suggests that surface relaxation and reconstruction play only a minor role in the separation process, indicating that the energetics are governed primarily by the intrinsic weak bonding across the van der Waals gap. These features, together with the weak bonding across the van der Waals gap, identify $Bi_4I_4$ as a viable candidate for the isolation of stable low-dimensional forms. In addition to thin layers, the pronounced structural anisotropy and weak inter-unit coupling suggest that narrowly confined ribbon-like and quasi-1D structures could be obtained under suitably controlled exfoliation conditions. Importantly, the near-identical cleavage energies of the two phases rule out a significant change in inter-unit binding strength as the origin of the observed Raman anomalies.

The phase transition occurs without a change in space-group symmetry, implying that the structural differences between the two phases are comparatively subtle. We therefore examined the atomic displacement patterns of the Raman modes of interest. The eigenvectors for the modes near 45, 55, 100, and 115 cm$^{-1}$ are shown in Figures 8 (a–d) for the α-phase and Figures 8 (e–h) for the β-phase. The calculated eigenvectors indicate that these Raman modes are mixed rather than purely shear- or breathing-type vibrations. Each mode involves a combination of Bi–Bi bond distortions and iodine motion, and the displacement patterns are not identical between the two stacking configurations, complicating a one-to-one qualitative classification. A common feature, however, is that these modes are predominantly polarized within the *ac* plane, with minimal displacement along the *b* direction[58]. This is consistent with the strong Raman response observed when the incident polarization is aligned along the *a*-axis. The stacking rearrangement modifies local bonding environments and consequently alters specific interatomic force constants in a mode-dependent manner. As a result, different phonons experience distinct renormalizations across the transition, leading to simultaneous hardening of some modes and softening of others[59].

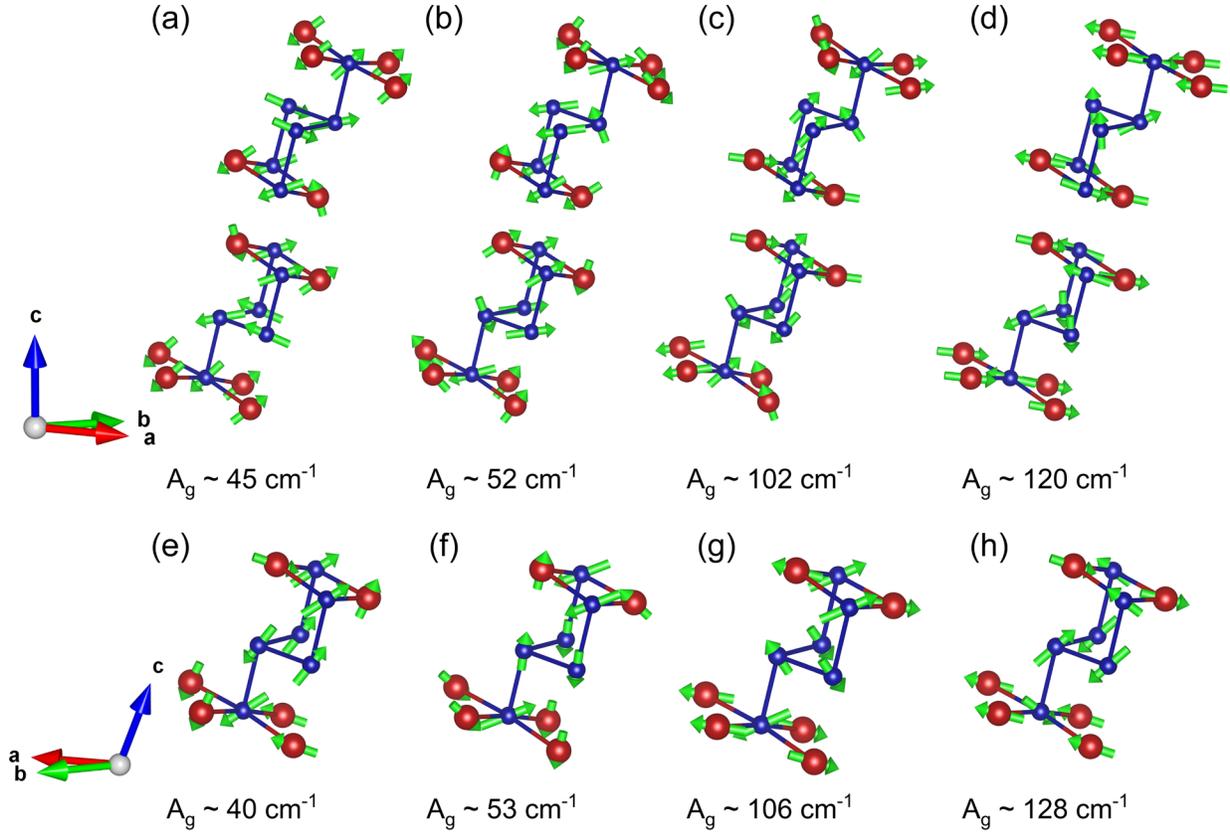

*[Fig. 8: Calculated atomic displacement patterns for the Raman-active modes of $Bi_4I_4$. (a–d) Eigenvectors of the α-phase for the modes at 45, 55, 100, and 115 $cm^{-1}$, respectively. For this phase, two neighboring chains within the same primitive cell along the c direction are shown to highlight their relative atomic displacements across the van der Waals gap. (e–h) Corresponding eigenvectors for the β-phase, shown for a single chain along c direction. In both phases, only part of the primitive cell is shown for clarity. Blue and red spheres denote Bi and I atoms, respectively; green arrows indicate the phonon displacement directions. The mode symmetry and the DFT-calculated frequency are labelled below each respective panel. The displacement patterns are shown only for a representative chain, while the crystallographic axes correspond to the primitive cell.]*

The structural transition that drives the change in topological band character deserves closer examination from a symmetry perspective. In the low-temperature α-phase, the surface states become gapped, and the system realizes a higher-order topological insulator[20]. Recent work has attributed this gapping to the acquisition of mass by the surface Dirac fermions through a mechanism involving spontaneous breaking of the equivalence between opposite edges of individual $Bi_4I_4$ monolayers upon the stacking rearrangement[60]. Crucially, however, the crystallographic space group (C2/m) and its centrosymmetric character are preserved across the transition; the point-group symmetry operations remain identical in both phases. What changes is the internal arrangement of inversion centers within the enlarged unit cell of the α-phase: the stacking rearrangement breaks the in-plane inversion symmetry within individual monolayers, rendering the two edges of each chain inequivalent, while the inter-layer inversion symmetry that defines the centrosymmetric space group is preserved. From a Raman spectroscopic standpoint, this situation is remarkable. In previously reported cases where Raman has been used to detect inversion-symmetry changes, most notably in $MoTe_2$, where new modes activated by the loss of centro-symmetry provided direct evidence for the non-centrosymmetric phase,[26] the space group itself changed, producing qualitatively new selection rules and the appearance of formerly forbidden modes. In $Bi_4I_4$, by contrast, the selection rules are nominally identical in both phases. The abrupt, hysteretic phonon anomalies documented here therefore represent a distinct category of Raman sensitivity: the detection of a local symmetry rearrangement within a globally preserved centrosymmetric framework, manifested not through mode activation but through mode-selective force-constant renormalization. To our knowledge, this type of spectroscopic detection has not been previously demonstrated in quasi-1D systems.

The characterization data presented on a topological phase transition occurring near room temperature may help open a pathway toward functional devices that can be switched using unconventional physical mechanisms. Because the transition modifies the topological band character through subtle changes in inter-chain stacking, it offers the possibility of reversible, low-energy control of electronic states, potentially enabling low-power logic or memory devices that exploit topological protection against backscattering. In addition, the quasi-1D nature and strong spin–orbit coupling in $Bi_4I_4$ make it a promising platform for applications, where the spin-momentum locking of topological surface states could be modulated across the phase transition, enabling electrically or thermally controlled spin transport. Furthermore, the pronounced polarization-dependent Raman response documented here reveals a strong optical anisotropy in $Bi_4I_4$, with the complex Raman tensor analysis indicating significant absorptive anisotropy along different crystallographic directions. This finding is directly relevant to the recently demonstrated potential of β-Bi$_4$I$_4$ as a broadband saturable absorber with modulation depths exceeding 60%[5], where polarization-dependent nonlinear optical response is a critical design parameter. The proximity of the transition temperature to ambient conditions further suggests opportunities for practical device integration. The demonstrated coupling between lattice dynamics and topological order implies that phonon engineering[61,62] could provide an additional degree of control over the material properties.

## 3. Conclusions

We investigated optical phonons in the quasi-1D material Bi$_4$I$_4$, which exhibits distinct topological insulating phases separated by a transition near room temperature. Using angle-resolved and temperature-dependent polarized Raman spectroscopy, we tracked the evolution of lattice dynamics across the α–β transition. Symmetry-resolved analysis enabled unambiguous assignment of the dominant $A_g$ and $B_g$ phonon modes, including identification of the 100 cm$^{-1}$ feature as an $A_g$ mode. The $A_g$ modes exhibit an anomalous angular dependence in the parallel-polarized configuration, arising from absorption-induced phase differences in the complex Raman tensor elements. Across the transition near 300 K, phonons display abrupt and reversible changes in frequency, intensity, and linewidth accompanied by a clear thermal hysteresis, establishing the first-order nature of the phase transition. The collective phonon anomalies reflect the stacking rearrangements of the Bi$_4$I$_4$ chains that occur without a change in crystallographic space group. Our results establish polarization-resolved Raman spectroscopy as a sensitive probe of stacking-registry-driven topological phase transitions that occur even in the absence of crystallographic symmetry breaking. This capability is important for further investigation of topological materials with transitions near room temperature for applications in future electronic and quantum technologies. More broadly, the combination of near-ambient topological switching, strong optical anisotropy, and the potential for laser-induced local phase control, positions Bi$_4$I$_4$ as a uniquely versatile platform at the intersection of topological electronics and anisotropic photonics.

4. **Experimental**

**Crystal growth and characterization of $Bi_4I_4$:** $Bi_4I_4$ crystals were grown using CVT from Bi and $HgI_2$ in a 2:1 molar ratio. The sealed growth ampule was maintained at a temperature gradient of 280-210 °C for one week before slow cooling. Full details can be found in the Supplementary Information. We observed excess $HgI_2$ as red particles on the surfaces of some $Bi_4I_4$ crystals in each run, as shown in Supplementary Figure S2. These were removed using hot acetone and brief bath sonication. We observed that this cleaning process causes partial fragmentation of the crystals, consistent with the weak interchain van der Waals interactions in $Bi_4I_4$ that enable cleavage along both the (001) and (100) planes[63].

Powder X-ray diffraction (PXRD) analysis of $Bi_4I_4$ proved to be challenging due to several factors that complicate phase identification: strong preferred orientation effects arising from the highly anisotropic quasi-1D structure, the close structural similarity between α- and β-phases, and the tendency of β-$Bi_4I_4$ to convert to the α-phase upon grinding[64]. These issues underscore the limitations of conventional PXRD for $Bi_4I_4$ phase discrimination and highlight the importance of single-crystal structure determination for unambiguous identification. For these reasons, we confirmed the structures of both α- and β-$Bi_4I_4$ by single crystal X-ray diffraction (SCXRD). Diffraction data were collected from the same crystal at 135 K and then at 302 K, allowing direct observation of the α to β transition within a single specimen (Supplementary Figure S3). We found β-$Bi_4I_4$ at 302 K, confirming that the CVT growth conditions yielded the expected phase, and observed α-$Bi_4I_4$ upon cooling to 135 K. The refined structures are in good agreement with previous determinations by von Schnering and coworkers[64]. For α-$Bi_4I_4$ at 135 K, the refined lattice parameters are $a = 14.1745(7)$ Å, $b = 4.4142(2)$ Å, $c = 19.8408(10)$ Å, and $\beta = 92.979(2)°$. For β-$Bi_4I_4$ at 302 K, the corresponding values are $a = 14.3592(9)$ Å, $b = 4.4229(3)$ Å, $c = 10.4811(6)$ Å, and $\beta = 107.885(2)°$, consistent with previously reported values[20,21]. Our structure of α-$Bi_4I_4$ at 135 K exhibits slightly contracted unit cell dimensions leading to a unit cell volume ~2% smaller than the reported RT structure[64].

**Differential scanning calorimetry measurements:** Differential scanning calorimetry (DSC) measurements were performed using a DSC 214 Polyma (NETZSCH-Gerätebau GmbH,

Germany). Approximately 16.1 mg of as-synthesized $Bi_4I_4$ was sealed in a standard aluminum crucible and measured over the temperature range of 250–330 K at a controlled heating and cooling rate of 5 K min$^{-1}$.

**Raman spectroscopy:** Polarization-dependent Raman measurements at various temperatures were performed using a Renishaw inVia micro-Raman system equipped with 633 nm and 488 nm excitation sources. The incident and scattered light were collected in a backscattering geometry through a 50× objective (numerical aperture = 0.50). A 2400 l mm$^{-1}$ grating was used for measurements with 488 nm excitation, while an 1800 l mm$^{-1}$ grating was employed for 633 nm excitation. Polarization analysis was achieved by placing linear polarizers in the collection path before the CCD detector to select parallel and cross configurations. For temperature-dependent Raman measurements, the samples were mounted in a Linkam cryostat, providing stable temperature control between 77 K and 398 K. To minimize laser-induced heating, the excitation power was maintained below 150 μW. Because $Bi_4I_4$ shows distinct transition temperatures upon cooling versus heating, we used 2 K temperature increments in the vicinity of the transition to resolve subtle changes in peak position. The temperature ramp rate was 2 K min$^{-1}$, and each spectrum was recorded only after the stage temperature stabilized to ensure completion of the phase transition[28].

**Density functional theory calculations:** First-principles calculations were performed within density functional theory using the projector-augmented wave (PAW)[65,66] method as implemented in the Vienna Ab initio Simulation Package (VASP)[67,68]. The exchange–correlation functional was treated within the generalized gradient approximation of Perdew–Burke–Ernzerhof (PBE)[69]. A plane-wave kinetic-energy cutoff of 520 eV was employed in all calculations. van der Waals interactions, which are essential for describing the interchain coupling in quasi-one-dimensional $Bi_4I_4$, were included using the Grimme D3 dispersion[70]. Both the α- and β-phases were fully relaxed, allowing lattice parameters, cell shape, and internal atomic positions to vary, until the total energy converged to $1\times10^{-9}$ eV and the residual Hellmann–Feynman forces on each atom were below $1\times10^{-4}$ eV Å$^{-1}$. Brillouin-zone integrations were carried out using uniform Γ-centered k-point meshes with a reciprocal-space resolution of approximately 0.02 Å$^{-1}$, corresponding to 12×12×5 for the β-phase and 12×12×3 for the α-phase, consistent with their distinct periodicities along the chain direction.

Phonon properties were subsequently computed using the finite-displacement (frozen-phonon) method as implemented in Phonopy[71] interfaced with VASP. For each relaxed phase, a 2×2×2 supercell was constructed from the optimized primitive structure. Symmetry-inequivalent atoms were displaced along Cartesian directions, and the resulting Hellmann–Feynman forces were calculated using single-point VASP calculations with the same 520 eV cutoff, PBE functional, D3 dispersion correction, and a stringent electronic convergence criterion of $1\times10^{-9}$ eV to ensure accurate force constants. The k-point meshes for the supercell calculations were reduced proportionally to the supercell size while maintaining equivalent reciprocal-space resolution. The interatomic force constants were extracted from the set of finite displacements and Fourier transformed to obtain the dynamical matrices and phonon frequencies throughout the Brillouin zone. The acoustic sum rule was enforced to eliminate residual translational drift at the Γ point. The absence of imaginary phonon modes confirmed the dynamical stability of both α- and β-$Bi_4I_4$ at the harmonic level.

**Acknowledgments**

The work at UCLA was supported, in part, by the Vannevar Bush Faculty Fellowship (VBFF) to A.A.B. under the Office of Naval Research (ONR) contract N00014-21-1-2947 on One-Dimensional Quantum Materials. The work at UCR and the University of Georgia was supported, in part, *via* the subcontracts of the ONR project N00014-21-1-2947. F.K. and A.A.B. also acknowledge the support of the National Science Foundation (NSF), Division of Materials Research (DMR) *via* the project No. 2205973 entitled "Controlling Electron, Magnon, and Phonon States in Quasi-2D Antiferromagnetic Semiconductors for Enabling Novel Device Functionalities." C.M. acknowledges support through the Quantum Networks Training and Research Alliance in the Southeast (QuaNTRASE), funded by the National Science Foundation through award No. 2152159. DFT calculations were performed on STAMPEDE3 at TACC and EXPANSE at SDSC under allocation DMR130081 from the Advanced Cyberinfrastructure

Coordination Ecosystem: Services & Support (ACCESS) program [38], which is supported by National Science Foundation Grants No. 2138259, No. 2138286, No. 2138307, No. 2137603, and No. 2138296. This work used the JEOL 2100PLUS microscope housed in UGA's Georgia Electron Microscopy core facility, acquired with funding from the National Institutes of Health through grant 1S10OD034282-01. The authors thank Dr. Pingrong Wei (UGA) for assistance with SCXRD analysis.

**Author Contributions**

**Competing Interests**

The authors declare no competing interests.

**The Data Availability Statement**

The data that support the findings of this study are available from the corresponding author upon reasonable request.